\documentclass[conference]{IEEEtran}
\IEEEoverridecommandlockouts
\usepackage{cite}
\usepackage{amsmath,amssymb,amsfonts}
\usepackage{algorithmic}
\usepackage{graphicx}
\usepackage{textcomp}
\usepackage{xcolor}
\usepackage{amsfonts}
\usepackage{booktabs}
\usepackage{float}
\usepackage{graphicx}
\usepackage{grffile}
\usepackage[algo2e]{algorithm2e}
\usepackage{longtable}
\usepackage{wrapfig}
\usepackage{gensymb}
\usepackage{textcomp}
\usepackage{amsmath}
\usepackage{algorithm}
\usepackage{graphicx}
\usepackage{caption}
\usepackage{listings}
\usepackage{array}
\usepackage{subcaption}
\usepackage{booktabs}
\usepackage{amsmath}
\usepackage{enumitem}
\usepackage{dblfloatfix}
\usepackage{lipsum}
\usepackage{soul}
\usepackage{fancyvrb}
\usepackage{xcolor}
\usepackage{verbatimbox,caption,float,lipsum}
\newfloat{Code}
\usepackage[utf8]{inputenc}

\title{Hierarchical Federated Learning based Anomaly Detection using Digital Twins for Smart Healthcare}

\newcommand{\linebreakand}{%
  \end{@IEEEauthorhalign}
  \hfill\mbox{}\par
  \mbox{}\hfill\begin{@IEEEauthorhalign}
}
\makeatother

\author{\IEEEauthorblockN{Deepti Gupta\IEEEauthorrefmark{1}, Olumide Kayode\IEEEauthorrefmark{2}, Smriti Bhatt\IEEEauthorrefmark{3}, Maanak Gupta\IEEEauthorrefmark{4}, Ali Saman Tosun\IEEEauthorrefmark{5} }
\IEEEauthorblockA{\IEEEauthorrefmark{1}Dept. of Computer Science,
University of Texas at San Antonio,
San Antonio, TX 78249, USA \\\IEEEauthorrefmark{2}Dept. of Computer Science, Frostburg State University, Frostburg, MD 21532, USA\\\IEEEauthorrefmark{3}Dept. of Computer and Information Technology, Purdue University,
West Lafayette, IN, 47907, USA\\\IEEEauthorrefmark{4}{Dept. of Computer Science},
Tennessee Technological University,
Cookeville, TN 38505, USA \\\IEEEauthorrefmark{5}Dept. of Math. and Comp. Sci., University of North Carolina at Pembroke, Pembroke, NC 28372, USA}
\IEEEauthorrefmark{1}deepti.mrt@gmail.com, \IEEEauthorrefmark{2}kofanilola@gmail.com, \IEEEauthorrefmark{3}smbhatt@purdue.edu, \IEEEauthorrefmark{4}mgupta@tntech.edu, \IEEEauthorrefmark{5}ali.tosun@uncp.edu}

\begin{document}

\maketitle

\begin{abstract}
Internet of Medical Things~(IoMT) is becoming ubiquitous with a proliferation of smart medical devices and applications used in smart hospitals, smart-home based care, and nursing homes. 
It utilizes smart medical devices and cloud computing services along with core Internet of Things (IoT) technologies to sense patients' vital body parameters, monitor health conditions and generate multivariate data to support just-in-time health services. Mostly, this large amount of data is analyzed in centralized servers.
Anomaly Detection (AD) in a centralized healthcare ecosystem is often plagued by significant delays in response time with high performance overhead. Moreover, there are inherent privacy issues associated with sending patients' personal health data to a centralized server, which may also introduce several security threats to the AD model, such as possibility of data poisoning. 
To overcome these issues with centralized AD models, here we propose a Federated Learning (FL) based AD model which utilizes edge cloudlets to run AD models locally without sharing patients' data. Since existing FL approaches perform aggregation on a single server which restricts the scope of FL, in this paper, we introduce a hierarchical FL that allows aggregation at different levels enabling multi-party collaboration.
We introduce a novel disease-based grouping mechanism where different AD models are grouped based on specific types of diseases.
Furthermore, we develop a new Federated Time Distributed (\textsc{FedTimeDis}) Long Short-Term Memory (LSTM) approach to train the AD model. 
We present a Remote Patient Monitoring (RPM) use case to demonstrate our model, and illustrate a proof-of-concept implementation using Digital Twin (DT) and edge cloudlets.
\end{abstract}
\begin{IEEEkeywords}
Federated Learning, Anomaly Detection, Internet of Medical Things, Remote Patient Monitoring, Security, Privacy, Edge Cloudlet Computing, Long Short-Term Memory, Digital Twin.
\end{IEEEkeywords}
\section{Introduction}
\label{Intro}
A prominent extension of Internet of Things (IoT) is the Internet of Medical Things (IoMT) that aims to improve the quality and accessibility of healthcare services. IoMT is enabling today's smart healthcare ecosystem using advanced technologies including 6\textsuperscript{th} generation~(6G) mobile networks, Cloud Computing, and Artificial Intelligence (AI). 
The burgeoning use of IoMT devices and data-driven applications assist in 
producing positive outcomes ranging from improved user health posture, early disease diagnosis, better quality treatment to cost effective smart healthcare ecosystem.
Majority of IoMT devices are wearable devices, such as smart watch, ECG monitors and other wireless devices such as smart oximeter, blood pressure monitor, glucose meter etc.
One of IoMT applications is Remote Patient Monitoring (RPM), which utilizes IoMT devices and existing technologies such as Machine Learning (ML), Artificial Intelligence (AI), Virtual Reality (VR) and Blockchain.

As predicted by Gartner\footnote{https://www.gartner.com/en/documents/3990045/forecast-analysis-healthcare-providers-iot-endpoint-elec} the amount spent on IoMT by healthcare providers will reach \$54 billion in 2029. As the number of smart medical devices increase, a tremendous amount of data is being collected from these devices which raises major security and privacy concerns in the IoMT domain. 
In this paper, we mainly focus on anomaly detection models to identify malicious users, devices, and data generated from these users or devices (e.g., user behavior patterns, device data for poisoning AD model, etc.).
In the past, various Anomaly Detection (AD) models~\cite{bi2016anomaly,deep2019survey, swaroop2019health, rachakonda2020ilog} have been developed to secure RPM ecosystem based on patients' behavior. Gupta et al.\cite{gupta2021detecting} proposed Hidden Markov Model (HMM) based AD for RPM by using smart home and smart health devices that analyzes anomalous users' behavior. 
More recently the US National Institute of Standards and Technology (NIST) published a report~\cite{cawthra1800securing} on the RPM ecosystem, which highlights possible security and privacy solutions to build a secure RPM infrastructure.

Despite advances in AD models for the RPM ecosystem, there are still growing concerns about safety, security and privacy of users. 
Therefore, this ecosystem requires a comprehensive and robust anomaly detection approach to detect anomalous behavior of various entities effectively. 
Generally, centralized AD models face many challenges in terms of security and privacy, such as patient data privacy, training data poisoning, model drift, and performance overhead as discussed in detail in the Section~\ref{threat}. 
In this paper, we raise the following research questions (RQs):


\begin{itemize}
\item RQ1 - While developing AD models for smart healthcare, how to enhance patients' data privacy?
\item RQ2 – How can anomaly detection models guarantee the integrity and statistical features of training data received from multiple smart devices?
\item RQ3 – What approaches can be used to reduce computation and communication cost in centralized server AD models? 
\item RQ4 - How distributed architecture for multi-user can enhance AD model accuracy compared to centralized anomaly detection approach? 
\end{itemize}

Considering the limitations of centralized AD models, our research enables us to obtain insightful answers to the above questions. As discussed in \cite{li2020federated, rieke2020future}, Federated Learning (FL) is a powerful tool that enables edge level training in various domains such as industrial IoT, smart home and healthcare. FL allows devices to collaborate with each other on the edge and perform training locally to build AD models without sharing user's data. 
This approach helps ensure data privacy and reduce device-to-cloud and cloud-to-device communication costs. Generally, FL utilizes a single server with multiple clients architecture, which may not be appropriate for certain computational needs. 
Thus, we propose a hierarchical FL that enables collaboration among various organizations using different levels of aggregation. For instance, if \textit{health organization-1} has only two diabetic patients, and \textit{health organization-2} has only one diabetic patient. In \textit{health organization-1}, both diabetic patients collaborate together to learn and build their local AD model. In \textit{health organization-2}, there is only one diabetic patient who needs to collaborate with other diabetic patients. Therefore, our hierarchical FL approach allows multiple healthcare organizations to collaborate at different levels for building robust AD models for patients with similar conditions. In this research, we develop a hierarchical FL based AD model to identify anomalies and secure the RPM ecosystem.

Our proposed approach utilizes Digital Twin~(DT) which is an advanced technology that helps to build a patient's digital representation. It allows the creation of a safe and secure environment where doctors can perform specific tests, such as testing any medication on a virtual patient without any risks. Using DT technology, physicians can provide customized medical treatments to patients. 
It can also provide a platform for enhanced interoperability among IoMT and smart home IoT devices from different vendors that may be jointly utilized. The amalgamation of DT and healthcare can provide more accurate and fast service delivery for RPM. In our model, DT allows participants to sink data with the model, which permits weight aggregation in a synchronous manner reducing wait time in FL.  


In this paper, we propose a Federated Time Distributed (\textsc{FedTimeDis}) Long Short-Term Memory (LSTM) 
framework to build local AD model for identifying anomalies using hierarchical federation employing DT and Edge Cloudlet Computing (ECC). 
The main contributions of this paper are as follows.
\begin{itemize}
    \item We identify specific potential threats in centralized anomaly detection (AD) models. 
    \item We design a hierarchical Federated Learning (FL) based AD model for smart health ecosystems.
    \item We propose Federated Time Distributed (\textsc{FedTimeDis}) LSTM approach to develop AD model.
    \item We demonstrate Remote Patient Monitoring (RPM) use case to identify anomalies using our proposed model and also present proposed implementation framework using Digital Twin (DT) and Edge Cloudlet Computing (ECC).
\end{itemize}

The remainder of this paper is organized as follows. Section~\ref{related} presents the literature review on AD models using machine learning approaches. The issues of centralized AD models and possible threats are discussed in Section~\ref{threat}. Section~\ref{proposed} presents a proposed hierarchical FL based anomaly detection model along with LSTM definition and a novel \textsc{FedTimeDis} LSTM framework. Section~\ref{Block} explains about the major building blocks to build our model. Section~\ref{usecase} presents the RPM use case and illustrates steps to develop the AD model in a federated setting. Conclusion and future work are discussed in Section~\ref{conclusion}.
\section{Related Work}
\label{related}
In this section, we discuss the related work on centralized and edge based AD models for a single user using various ML approaches. 
Zhang et al.~\cite{zhang2013medmon} proposed MedMon, which is a medical security monitor that identifies anomalies in medical implantable devices. In \cite{deep2019survey}, researchers presented a survey on AD models and identified a suitable AD approach that uses dense-sensing networks for elderly people. 
Peddoju et al. \cite{peddoju2019health} presented an AD model for RPM at the device level by developing the ML model on the edge devices. 
Lee et al. \cite{lee2020rere} introduced real-time ready-to-go proactive AD model using LSTM model to predict anomalous scenarios without requiring human intervention. Fang et al.~\cite{fang2020practical} proposed an AD model to identify illegal behavior in the medical IoT environment based on rough set theory and fuzzy core vector machine (FCVM). Ren et al.~\cite{ren2017anomaly} presented a dynamic Markov based AD model, where patterns of training data are managed by sliding windows.  
Home Automation Watcher (HAWatcher) \cite{fu2021hawatcher} is a semantics-aware AD system for appified smart homes.
In addition, several security models for protecting IoT devices are discussed in~\cite{gupta2020access, kayode2020towards, chaganti2021intelligent, gupta2020learner, gupta2021game, gupta2021future, aslan2021intelligent, ozkan2021comprehensive, bhatt2017access2, bhatt2021attribute}.

Fatim et al. \cite{fahim2019anomaly} conducted a literature review about AD models and identified several issues including imbalanced large datasets, limitations of ML approaches, and lower accuracy in identifying abnormal behavior in real-world scenarios. In \cite{said2021efficient}, AD and intrusion detection systems are proposed for smart hospitals to detect rare events with respect to patient's health. These models are deployed on the edge. Schneble et al. \cite{schneble2019attack} introduced ML based intrusion detection model for medical Cyber-Physical Systems and explored FL to reduce the communication and computation costs. Sater et al. \cite{sater2020federated} presented an AD model in a FL setting to solve multiple issues simultaneously in a smart building. Kiranyaz et al. \cite{kiranyaz2015real} introduced a patient-specific ECG heartbeat classifier with an adaptive implementation of 1-D CNNs.

Prior research has introduced various AD models for medical and smart home domains, and these models have been deployed either on centralized servers, or edge devices for a single user. These models have used various ML approaches to classify the data into normal/abnormal categories. Gupta et al. \cite{gupta2021detecting} introduced a centralized AD model for a single user in the RPM ecosystem. However, as discussed earlier, centralized AD models are facing privacy, high latency and high communication cost issues. On the other hand, edge device level models are facing low accuracy rate to identify anomalies due to low volume of data. A robust AD model is still lacking, which can provide the security and privacy solutions for protecting patient's health data. 

To the best of our knowledge, a federated time distributed based LSTM model has not been used in healthcare domain to identify anomalies in a federated setting. 
To bridge this gap, we believe our proposed hierarchical FL based AD model offers a novel perspective to detect anomalies in the smart healthcare ecosystem. 

\begin{figure}[t]
\centering
\includegraphics[width=0.5\textwidth, height=.16\textheight]{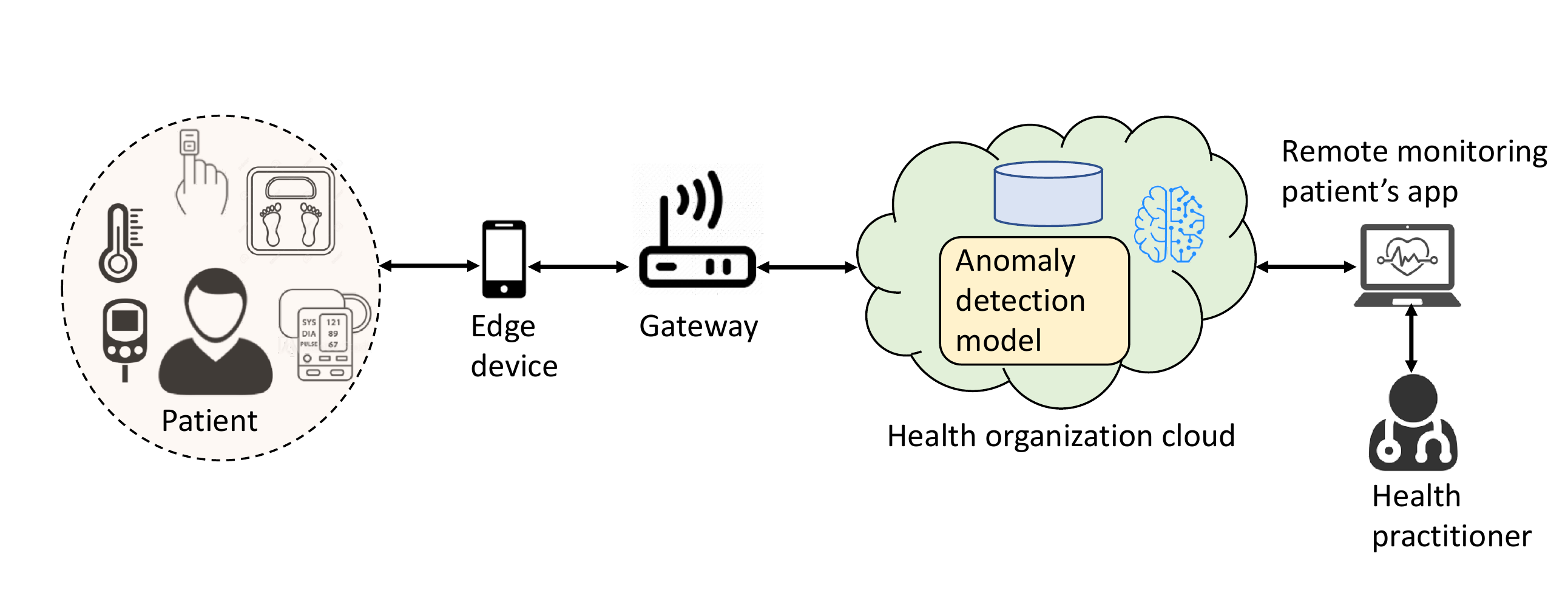}
\centering
\caption{Centralized Anomaly Detection Model for RPM.}
\label{fig:center}
\end{figure}
\section{Threat Model}
\label{threat}
In this section, we discuss specific threat scenarios associated with centralized AD models. 
These threat scenarios motivated this research and led to the development of our hierarchical Federated Learning AD model.

\subsection{Centralized Anomaly Detection Model}
A centralized data processing approach for AD models has become popular due to its advantages such as ease of data access, common data repository, simplicity, and fast read/write operations. 
Figure \ref{fig:center} depicts a centralized AD model for RPM which utilizes a centralized data repository hosted in the health organization cloud. 
In this architecture, the AD model and dataset are close together which enables ease of model training and better read/write performance.

\begin{figure*}[t]
\centering
\includegraphics[width=1\textwidth, height=.37\textheight]{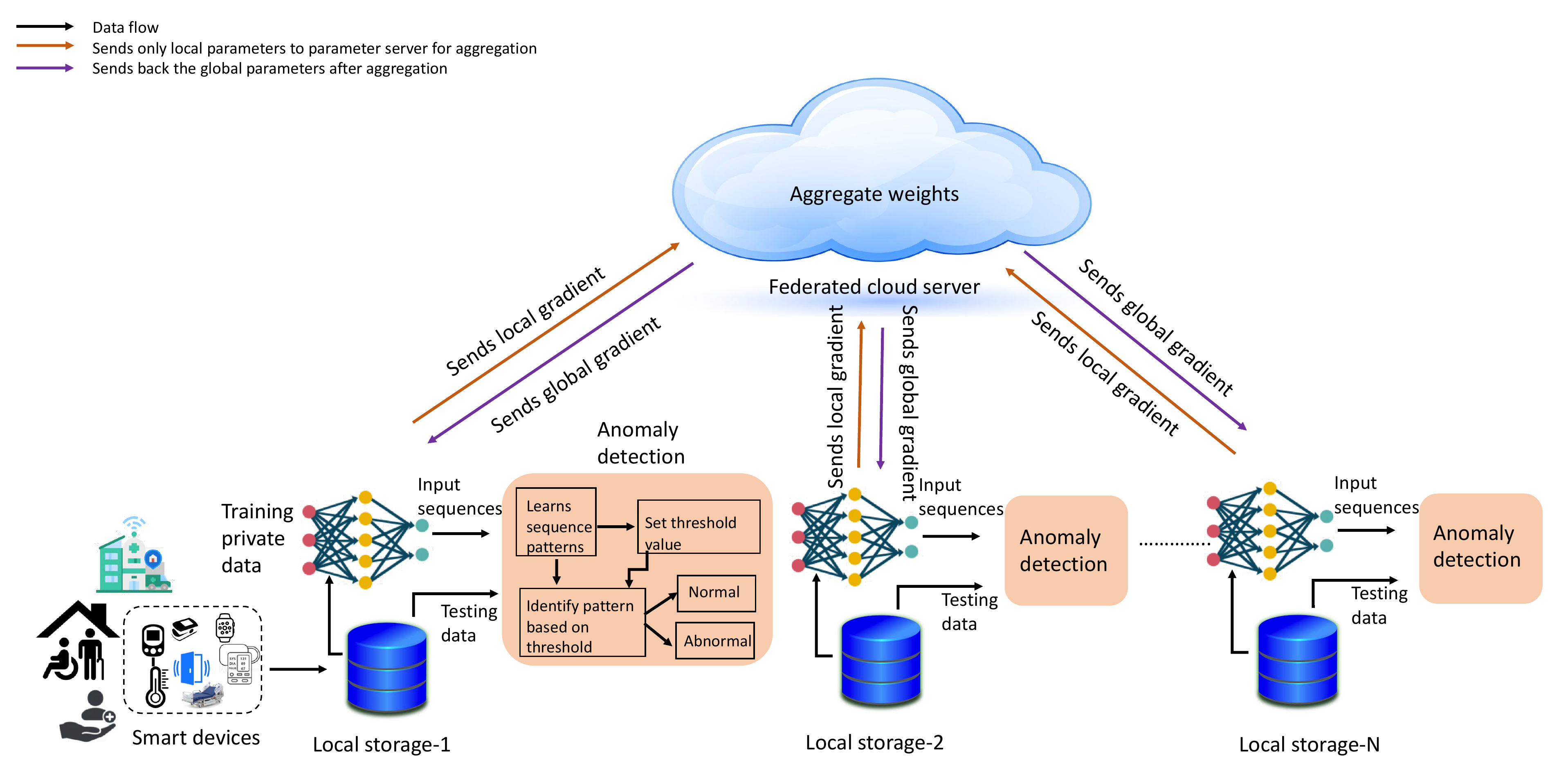}
\centering
\caption{Federated Learning based Anomaly Detection Model.}
\label{fig:Model}
\end{figure*}
However, centralized data processing and AD approach have some inherent limitations. One of the prominent weaknesses of centralized servers is a single point of failure that can hinder the availability of model, data or services. Another major concern is data privacy since data from multiple sources (e.g., users) is stored at a central location. In a centralized data repository, one client may be able to access or modify other clients' data if appropriate access control and segregation mechanisms are not implemented. In such a case, the data from multiple sources could be corrupted due to intentional or accidental modifications. 
Moreover, there are several other issues, such as node failure, network communication overhead and latency, data interruption due to network congestion/bottleneck, that can degrade the system performance limiting the suitability of centralized AD approach for some critical application domains as healthcare.  

\subsection{Threat Scenarios}
Here, we discuss some of the possible threats associated with the centralized AD approach. We aim to address these threats using FL.

\subsubsection{Privacy Leakage}
Data privacy is a key concern for users in any data driven application, especially in the healthcare domain. Some important questions that need to be addressed in this domain are -- \textit{Who has access to what set of data? How are the data and its attributes being used by other nodes or clients within a network?} In a centralized data processing approach, malicious users may use advanced techniques to compromise the system and gain access to sensitive health information pertaining to any individual. It is necessary to design and deploy novel mechanisms to ensure privacy of patients considering that a model may reveal certain aspects of patients' data and information. At the same time, patients also need to have adequate transparency and control of their data. Using FL and a grouping mechanism (based on specific diseases and patient attributes), we can ensure that only local AD model assigned to a patient can have access to their sensitive information without sharing it with other AD models belonging to different patients. 

\subsubsection{Training Data Poisoning}
Data poisoning in ML entails polluting a model's training data. It is generally considered an integrity attack because tampering with the training data impacts the model's ability to correctly detect anomalies or output the right predictions. In a centralized approach, where all the training data is exposed to a model, an attacker can introduce data samples that can decrease the accuracy of that model, so that eventually outliers, strange values or falsified data won't be flagged as anomalous. Using FL where training data are stored locally on each client, it will be difficult to poison such localized data given security of the client is not compromised. For example, data from diabetic patients are not mixed with data from sleep apnea patients. Securing communication channel and data flow between patients and their corresponding digital twins as we plan to demonstrate in our proposed approach will greatly protect our neural network model from training data tampering. 

\subsubsection{Model Drift}
A fully trained neural network model may deteriorate after some time due to changes in new data or falsified data. Generally, a model experiences a drift when the underlying statistical structure of the data changes over time. Attackers can introduce new data points that can alter the inherent statistical structure of the training data when vulnerabilities in the architectural design of the framework or data storage are exploited. Also, different units of measurements and ranges of valid data values for different diseases can lead to a problem for the model. Using FL and disease-based grouping, we can ensure that data generated from patient devices are expected values that will not to a great extent alter or adversely affect the statistical structure of patient's data. 

\subsubsection{Performance Overhead}
There are several factors that can affect the performance of AD models, such as memory constraint, latency, system failure to packet loss along the communication channel. In a centralized approach, a single point of failure of the repository can affect the availability of data and render the neural network model ineffective. Overhead due to loss of data or high response time are other major issues that can affect the performance of an AD model. We can address some of these issues using FL to mitigate bottlenecks during data access. In addition, performing decentralized training on each client's device or edge cloudlet will greatly improve the response time. 
Centralized data processing systems require transferring huge patient's data for every round of training which requires more computational resources compared to just uploading local or downloading global gradients from server and also would use minimal network parameters. 

To address these threat scenarios in a centralized AD model, we propose a hierarchical FL based AD framework with edge computing capabilities.


\section{Proposed Model}
\label{proposed}
In this section, we present our hierarchical FL based AD model for the smart healthcare ecosystem. 

\subsection{Federated Learning based Anomaly Detection Model}
\label{SubFL}

A smart connected healthcare domain consists of various components, such as IoT devices, different types of users (e.g., patients, physicians, etc.), edge gateway devices, cloud computing services, and data-driven applications. Within this complex ecosystem, there could be several anomalies arising due to various reasons, such as faulty IoT devices, critical conditions of a user, or malicious users in the system, etc. Therefore, we need a secure and robust anomaly detection model that can also address the threats associated with centralized AD models, as discussed in Section \ref{threat}. Here, we propose a privacy-preserving AD model using Federated Time Distributed (\textsc{FedTimeDis}) LSTM for a connected healthcare ecosystem. 


Figure \ref{fig:Model} depicts our FL based AD model. Here, \textit{N} is the number of participants (e.g., patients, smart hospitals) and depends on specific use case requirements. We assume each participant is connected with \textit{K} number of smart IoT devices and has its own local storage dataset $D_i$, where $i=1,2,3,$\ldots$,N $, collected through \textit{K} smart devices. 
Generally, smart devices generate a huge amount of data which is used to train the ML model, however, it is exposed to several privacy and other major issues. 
To overcome these issues, ML based AD models can train the data locally in a federated setting without compromising the patient's privacy. 
Before the start of training in a federated setting, 
we consider each participant signs an agreement to prove itself as legitimate participant. 
Then, each participant starts training to build a local model and maintains a local vector of neural-network parameters $w^i$, based on the training dataset. First, each local model initializes weights $w^i$  where $i=1,2,3,$\ldots$,N $, and sends them to the federated cloud server.
This server receives all the weights from all the participants' local models, and then aggregates them based on their attributes (e.g., user's age, disease name)
and sends them back as a global weight ($w_{globalz}$), as shown in the Algorithms \ref{algo:FTDL} and \ref{algo:group} (discussed in Subsection \ref{FedTimeDisLSTM}). 

In our model, the training data $(X^N,y^N)$ is defined for $N$ number of participants, each participant generates these time series input sequences based on devices' activation/deactivation. We also consider that many smart devices can activate simultaneously. For instance, a patient's wearable glucose meter and smart watch can send data simultaneously. The $y^N$ is the output of the model, which belongs to 0 and 1 to classify ``normal" and ``abnormal" observations respectively. During the training phase, the input is reconstructed in the output until reconstruction error is minimized, which identifies a threshold value. This threshold value decides the user's health, a device's status and his/her behavior is ``normal" or ``abnormal". Now, we discuss the LSTM approach and \textsc{FedTimeDis} LSTM framework.

\begin{figure}[t]
\centerline{\includegraphics[scale=0.4]{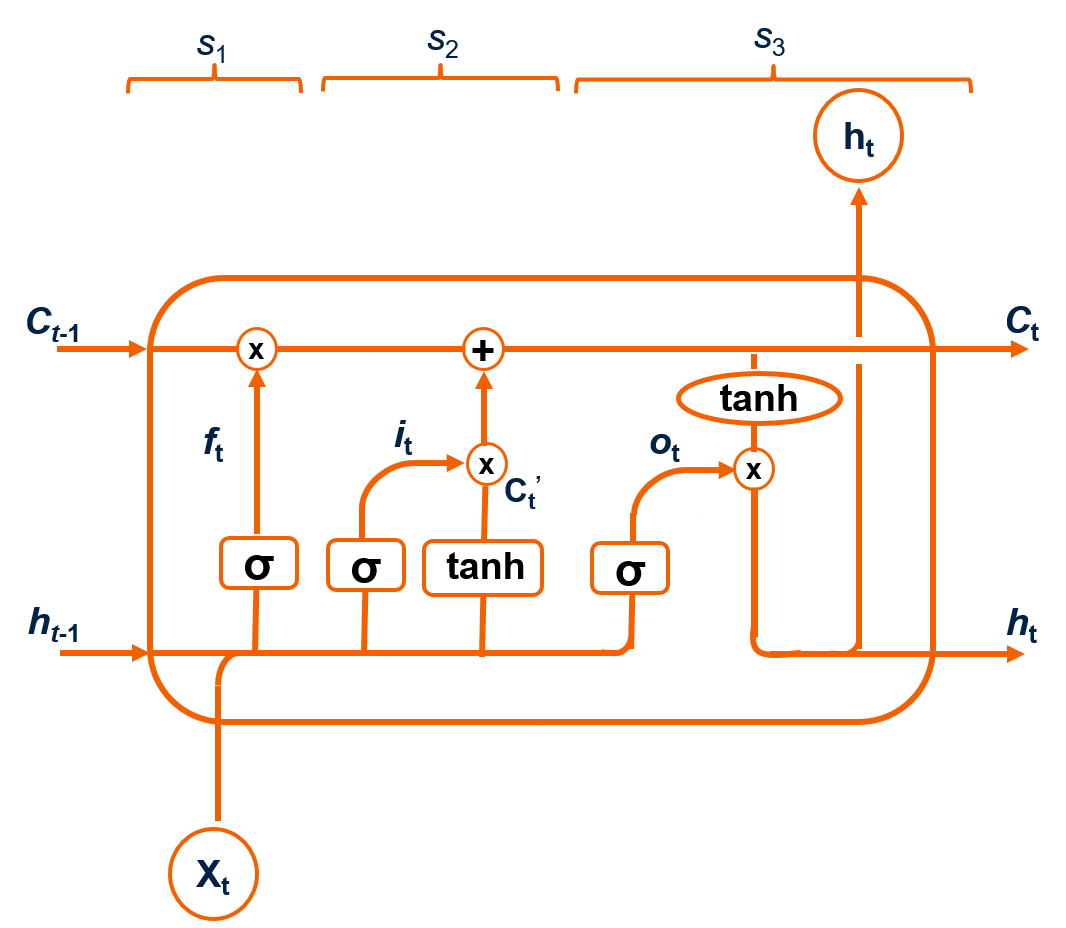}}
\caption{LSTM Cell}
\label{fig3}
\end{figure}

\begin{figure*}[t]
\centering
\includegraphics[width=1.0\textwidth, height=.41\textheight]{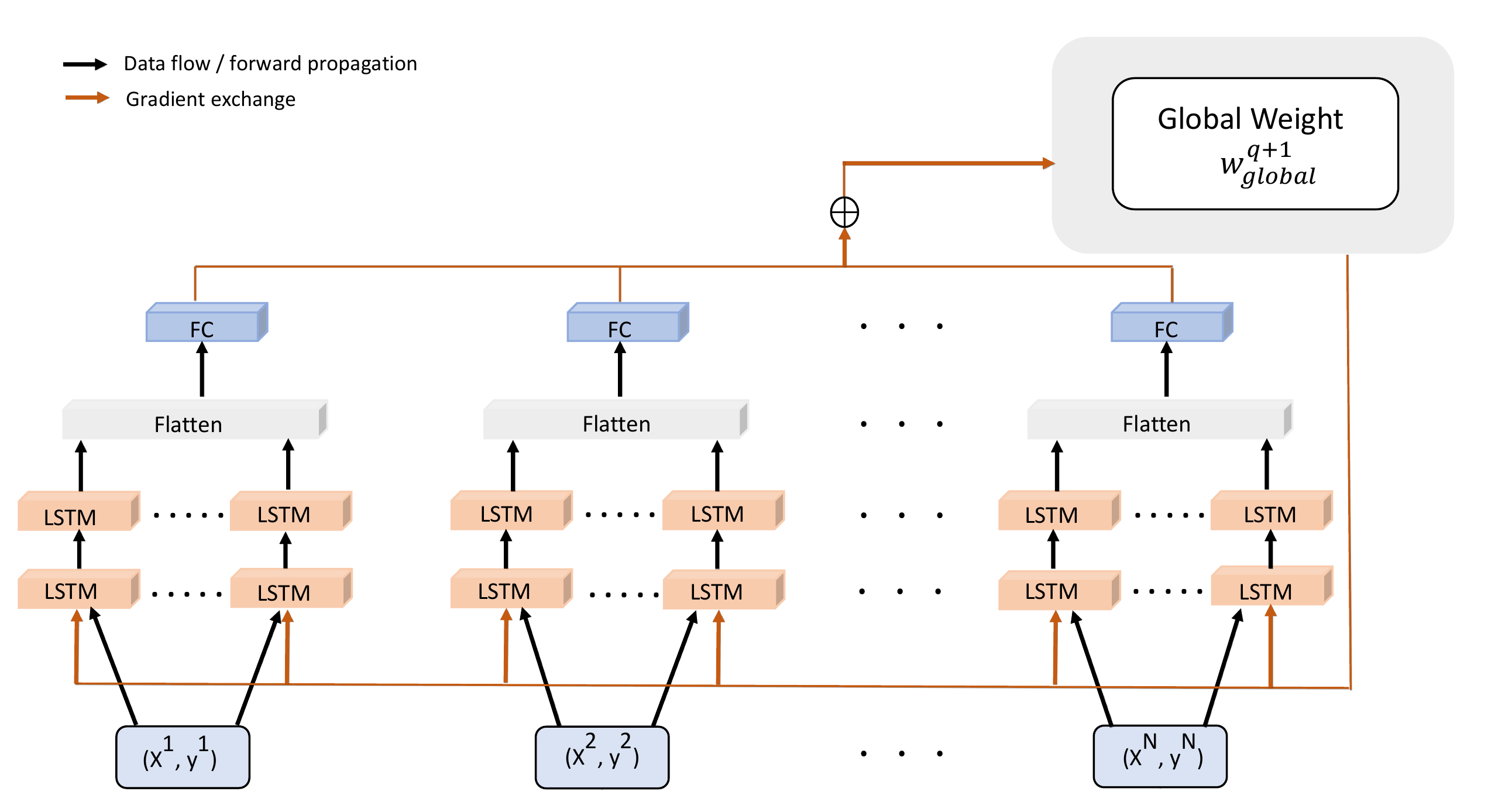}
\centering
\caption{Architecture of Proposed Federated Time Distributed LSTM Framework.}
\label{fig:FDL}
\end{figure*}

\subsection{Long Short-Term Memory (LSTM)}
\label{LSTM}
There are many types of neural networks used for learning applications of AI to solve real world problems. Regardless of a domain and nature of available data, intelligent solutions based on neural network models can be used for classification, regression, or AD. Different networks ranging from simple ones like Multilayer perceptron, Feed forward, to advanced ones like Boltzmann machine, Recurrent Neural Network (RNN) and LSTM~\cite{hochreiter1997long} have been used extensively in the literature. LSTM is a kind of RNN that is suitable for tasks associated with learning temporal dependencies in sequential data. It aptly captures time dependencies inherent in data and does not suffer from the vanishing or exploding gradient problem that is common in RNN. The LSTM network utilizes a memory cell and each cell is used to remember values over arbitrary time intervals depending on the architectural layout of the model.

Generally, a LSTM unit has three logistic gates that are used to regulate the flow of information in and out of the cell. As shown in Figure \ref{fig3}, the forget \emph{$f_{t}$} gate determines the information to keep or delete in the cell. The input \emph{$i_{t}$} and output \emph{$o_{t}$} gates are used for read and write respectively. At stage $s_{1}$ and at time \emph{$t$}, the previous cell state $c_{t-1}$ and output $h_{t-1}$ are fed into the current cell along with the current input data $x_{t}$ to decide what information to forget. At the second stage $s_{2}$, using the input gate, the current cell state $c_{t}$ is updated with new information. After stage $s_{2}$, the $c_{t}$ would have been updated twice, first to decide what aspect of the information from the previous cell to forget or keep, and second whether to add new information within the current cell or not. The last stage $s_{3}$ helps to compute the output using the \emph{$o_{t}$} gate to determine the information to be fed into the next unit. Across all units within the LSTM, both hidden and cell states are updated and transmitted to keep track of the long term dependencies in sequence of input data. Specifically, the implementation of LSTM cells can be illustrated using the following equations:
\[
\emph{$f_{t}$} = \sigma(\emph{$W_{f}$}[\emph{$h_{t-1}$},\emph{$x_{t}$}] + \emph{$b_{f}$}),        \tag{1}
\]
\[
\emph{$i_{t}$} = \sigma(\emph{$W_{i}$}[\emph{$h_{t-1}$},\emph{$x_{t}$}] + \emph{$b_{i}$}),        \tag{2}
\]
\[
\emph{$C_{t}^{'}$} = tanh(\emph{$W_{c}$}[\emph{$h_{t-1}$},\emph{$x_{t}$}] + \emph{$b_{c}$}),      \tag{3}
\]
\[
\emph{$o_{t}$} = \sigma(\emph{$W_{o}$}[\emph{$h_{t-1}$},\emph{$x_{t}$}] + \emph{$b_{o}$}),      \tag{4}
\]
\[
\emph{$C_{t}$} = \emph{$f_{t}$} \text{*} \emph{$C_{t-1}$} + \emph{$i_{t}$} \text{*} \emph{$C_{t}^{'}$},     \tag{5}
\]
\[
\emph{$h_{t}$} = \emph{$o_{t}$} \text{*} tanh(\emph{$C_{t}$})   \tag{6}
\]
where the weights and biases to be computed during the learning process are \emph{$W_{f}$}, \emph{$W_{i}$}, \emph{$W_{c}$}, \emph{$W_{o}$}, \emph{$b_{f}$}, \emph{$b_{i}$}, \emph{$b_{c}$} and \emph{$b_{o}$}. 

The operation \** represents element-wise multiplication of two vectors. The $\sigma$ is element-wise logistic sigmoid activation function while \emph{tanh} is element-wise hyperbolic tangent activation function. The former activation function produces output that is always between 0 and 1 while the latter activation function produces output within the range -1 and 1. The equations 3 and 5 show the two updates on cell state, and equation 6 shows that the hidden state is a function of the cell state and current output. Based on the aforementioned implementation details, the hidden state can also be referred to as short term memory while the cell state can be referred to as long term memory. We utilize all functionalities of the LSTM in our proposed model for AD.

\subsection{Federated Time Distributed LSTM Model}
\label{FedTimeDisLSTM}

Most models with a single unit or layer of LSTM can only accept one input sample per time and process it for sequential learning. Although we can learn inherent variability exhibited by the features in data using such conventional settings, the model may not perform well in capturing the sequential progression of changes in state. This is more evident in video data consisting of many frames. Stacks of frames per time will be better in showing transition of events than a single frame. The stacked LSTM, also known as deep LSTM, was first formulated by \cite{graves2013generating} and was applied to speech recognition problems. Exposing a model to multiple frames per time will enhance the extent of temporal dependencies learning. Processing multiple input data samples per time in succession will invariably help in prompt detection of anomaly across data generated by several IoT devices.

We propose a \textsc{FedTimeDis} LSTM approach for AD in the smart healthcare ecosystem. This approach allows us to pass multiple input samples to the model simultaneously for many participants. Additionally, Figure \ref{fig:FDL} shows that different sets of training data $(X^{N}, y^{N})$ from each participant are passed to a copy of the model running in a distributed processing manner. A time distributed model can use any number of neural network units and layers for optimal performance. For each layer, \emph{l} number of cells can be arranged sequentially and in \emph{m} stack. Depending on the value of \emph{l}, an equal number of input data samples are sent at the same time for processing. Our proposed architecture utilized four LSTM cells and two stacked layers before the Fully Connected (FC) or dense layer. The reason for our chosen value of \emph{l} and \emph{m} is to ensure fast processing, less computational resources requirement and optimal solution with better performance.  

The learning algorithm at the object layer is replicated across multiple IoMT users in several healthcare organizations. 
We initialize each weight parameter with an arbitrary value and set the number of epochs to \emph{H}. For local training, we use \textsc{FedTimeDis} LSTM, which is explained in the Algorithms \ref{algo:FTDL} and \ref{algo:group}. Here, a sequence of four input samples (as per use case requirement) are sent to the neural networks for each round. Hence, for a scenario where data are generated in real-time at a predefined interval, the model is not fed with any input samples until a collection of four data samples are available. 

\begin{algorithm}[H]
\SetAlgoLined
\caption{Federated Time Distributed LSTM}
\label{algo:FTDL}
\begin{algorithmic}[1]
\STATE{\textsc{FedTimeDis} LSTM $(X^N,y^N)$}
\STATE{Take inputs training sets $(X^n,y^n)$, where $n= 1,2,$\dots$,N$ from $N$ participants, initial global model parameter $w_{globalz}$, local minibatch size $J$, number of local epochs $H$, learning rate $\alpha$, number of rounds $Q$, $h$ hidden layer.}
\FOR{$q$ = 1 to $Q$}
\STATE{Federated cloud server randomly selects a dataset $D_q$ from $N$ participants.}
\STATE{Federated cloud server broadcasts $w_{globalz}^q$ to the participants using GROUPING algorithm.}
\FOR{each participant n $\in$ $D_q$ in simultaneously}
\STATE{${w_{globalz}^{q+1,n}}$ $\leftarrow$ update\_local(n, $w_{globalz}^q$)}
\ENDFOR
\STATE{${w_{globalz}^{q+1}}$ $\leftarrow$ ${\mathlarger{\sum}}_{n=1}^{N}$$\frac{D_n}{D}$${w_{globalz}^{q+1,  n}}$}
\ENDFOR \\
\textbf{update\_local($n$, $w_{globalz}^q$)}
\STATE{Split local dataset $D_i$ to minibatches of size $J$ which are included into the set $J_i$ and fed horizontally to four LSTM cells.}

\FOR{each local epoch $j$ from 1 to $H$}
\FOR{batch $(X, y)$ $\in$ $J$}
\STATE${h\textsubscript{t} = LSTM(h\textsubscript{t-1}, x\textsubscript{t}, w\textsuperscript{n})}$
\STATE{$y$\textsuperscript{n} = ${\sigma(W\textsuperscript{FC}h\textsubscript{2nd}+Bias)}$}\\
\STATE{$u$\textsuperscript{n} = $w$\textsuperscript{n} - ${w_{globalz}^n}$ }\\
\STATE{${w_{globalz}^n}$ $\leftarrow$ ${w_{globalz}^n}$  + $\frac{\alpha}{N}$ ${\mathlarger{\sum}}\textsubscript{n $\in$ $D_i$} u\textsuperscript{n}$}
\ENDFOR
\ENDFOR 
\STATE{return final weights $w_{globalz}^n$ to local edge server and start training again until minimizing the error to build the local anomaly detection model.}
\end{algorithmic} 
\end{algorithm}

\begin{algorithm}[H]
\SetAlgoLined
\caption{GROUPING algorithm}
\label{algo:group}
\begin{algorithmic}[1]
\STATE{Set initial global parameters $w_{global}$ based on disease types and patient's attributes (e.g., age, health conditions.}
\STATE{Builds the initial z groups using classification algorithm, now each group has own parameters $w_{globalz}$.}
\STATE{Sends the global parameters $w_{globalz}$ to the corresponding group participants}
\STATE{Federated cloud server aggregates the received local gradients in its assigned groups z using attributes synchronously.}  
 \end{algorithmic} 
 \end{algorithm}

For a situation where we have all the dataset available from different IoT devices for a patient, we pass a sequence of four samples $s_i$ to $s_{i+3}$ from local dataset $D_i$ for \emph{i} with a starting index of 0. For the next iteration, data samples $s_{i+4}$ to $s_{i+7}$ are passed to the neural network model. Hence, the samples $s_i$ to $s_{i+r}$ where \emph{r=3} are utilized for every iteration in our work. Local training on each device is carried out until error is below a specified minimum value or \emph{H} of epochs is obtained.
Subsequently, the parameters from all local models are uploaded to the server for aggregation. The federated server aggregates the model parameters submitted by the edge cloudlet. The actual learning phase involves training of the local model on edge cloudlets and sends back global parameters to each participant. 
The Algorithm \ref{algo:group} illustrates steps on how global parameters aggregate based on the patients' attributes and build groups for enhancing the accuracy of AD. After receiving the global parameters $w_{globalz}$, each device or edge cloudlet completes the training. This whole process of local training at edge cloudlet and gradient aggregation at the federated cloud server continues for all rounds until local models are fully optimized to effectively detect anomalies using the best possible values for the weights. Since the actual learning phase involves training of the local model at edge cloudlet, anomaly detection will be generated at edge. Moreover, the learning process of our model is real-time since the patient's data is continuously being generated from IoMT devices and fed into the model.

\section{Building Blocks for Proposed  Model}
\label{Block}
In this section, we describe the major building blocks for our proposed model. 

\begin{figure*}[t]
\centering
\includegraphics[width=1\textwidth]{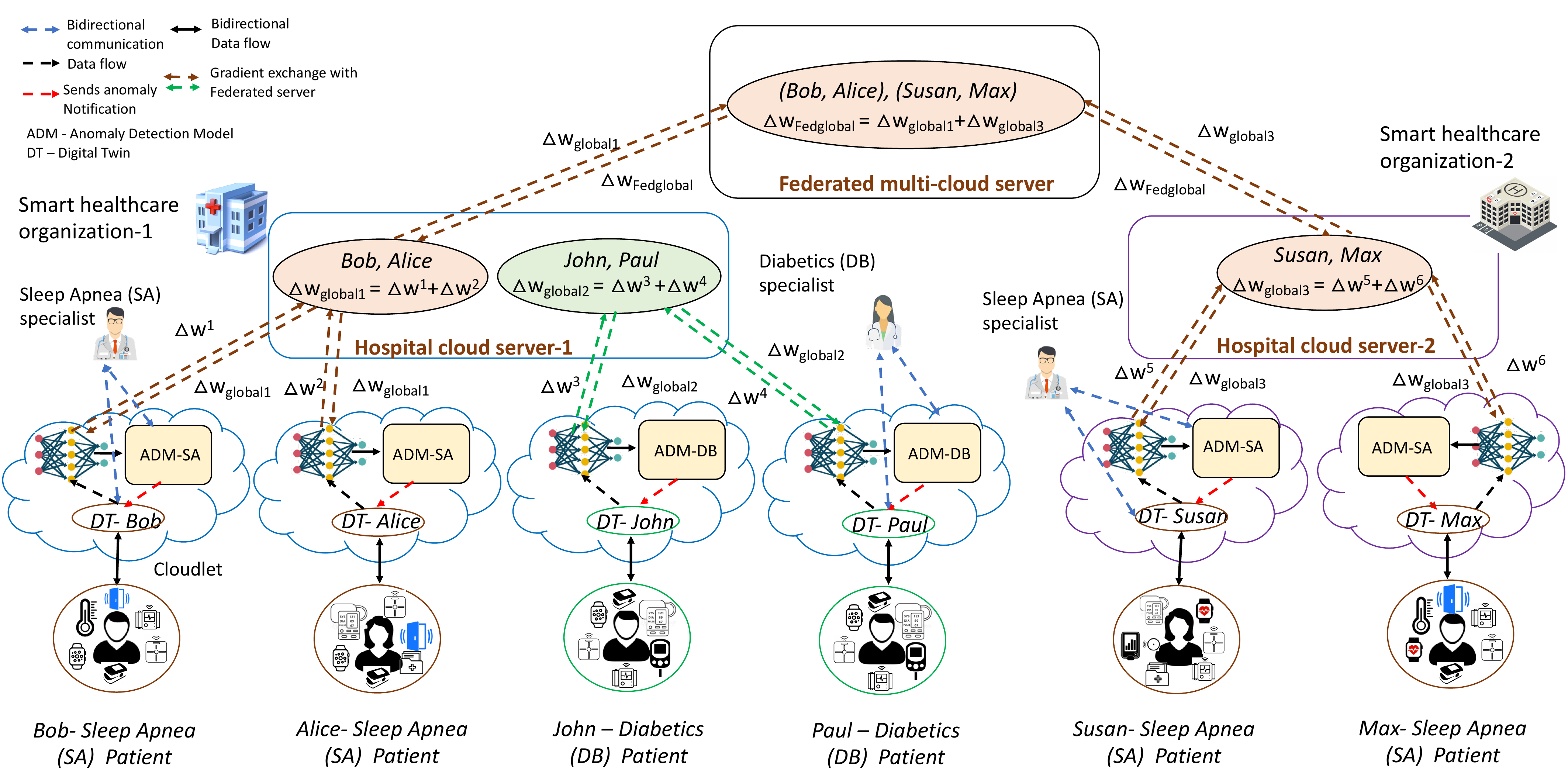}
\centering
\caption{A Use Case of Remote Patient Monitoring.}
\label{fig:usecase}
\end{figure*} 

\subsection{Digital Twin (DT)}
\label{DT}
DT is a dynamic virtual representation of a patient, which receives data from IoMT devices (wearable and non-wearable) and surrounding environmental sensors, and also includes patient's health records. 
The physical entity (e.g., patient) and its DT are intertwined with each other as they both exchange data continuously. In the healthcare domain, a DT provides an accurate real-time status of the patient and also enables health practitioners to provide specific services and test an optimal solution for improving the health of patients. Furthermore, DT allows to enhance security of the IoT integrated smart healthcare ecosystem.

In a federated setting, healthcare professionals collaborate effectively to learn about a patient's health through DT, and also prescribe right treatments, test these treatments through DT. It also addresses the issue of interoperability in the RPM ecosystem which consists of heterogeneous devices with different communication protocols and manufacturers. With the usage of DT, a federated server is able to aggregate the weights synchronously under stable data communication. 

In this paper, we propose the use of DTs on edge to reduce the gap between physical objects and their digital representations which are generally hosted in the cloud servers. Current research work \cite{castellani2020real,lu2020digital,xu2021digital} present the possibility of training data for AD model on DT in industrial IoT domain. However, in the healthcare domain, patient data privacy is a critical concern and we cannot merge one patient's data with another patient's DT, so training of datasets cannot be performed on a patient's DT in a collaborative manner. 
In this research, we design FL based AD model and deploy this model on edge cloudlet associated with the patient's DT. We provide the details of our implementation in Section \ref{usecase}. 

\subsection{Edge Cloudlet Computing (ECC)}
\label{CC}
In the past years, cloud computing has been considered the de-facto solution for IoT architectures. 
Due to various issues associated with central cloud architecture including privacy of data, delay in response time, low bandwidth and high latency, and high communication costs, Satyanarayanan et. al \cite{satyanarayanan2009case} introduced the concept
of virtual machine based edge cloudlets. Our proposed FL based AD framework utilizes edge cloudlet computing (ECC) to ensure data privacy and better model performance. 
Prior research \cite{bhatt2017access,bhatt2020abac} have proposed use of cloudlets for edge computing in RPM and healthcare domain.
ECC allows affordable local training of the model, especially in terms of costs and latency. From the implementation aspect, we consider that a Raspberry-Pi or its clusters can be used as ECC nodes where we can create enough computing power to deploy DT and train the local model.
\section{Use Case and Proof-of-Concept}
\label{usecase}
RPM is unfolding in many ways where patients receive better healthcare services at their homes, and health practitioners can access tangible AI-driven insights to handle the patient's health remotely, especially relevant in today's COVID-19 era. 
Figure \ref{fig:usecase} shows the RPM use case for multiple users utilizing proposed \textsc{FedTimeDis} LSTM approach, DT technology and ECC to identify anomalies in a federated setting.

\subsection{Remote Patient Monitoring (RPM)}
This proposed RPM use case presents a real-world scenario where patients are being monitored by health practitioners continuously. We consider that there are two smart healthcare organizations. \textit{Bob, Alice, John and Paul} belong to \textit{smart healthcare organization-1}, and \textit{Susan and Max} belong to \textit{smart healthcare organization-2}. \textit{Bob, Alice, Susan and Max}, who are of age group 30-36, have been diagnosed with \textit{Obstructive Sleep Apnea (OSA)} disease. On the other hand, \textit{John and Paul}, who are of age group 34-40, have been diagnosed as \textit{Diabetics (DB)}. These patients are being remotely monitored by their assigned \textit{OSA/DB specialist} continuously through patient's DTs and IoT applications.
Patients' health/activities generated data is captured by IoT devices and is sent to the DT service that builds a DT, which is digital representation of the patient. In the RPM environment, the assigned health practitioners have access to their patient's data through DT and will receive alerts if any anomaly occurs through AD model. This proposed AD model trains the data using \textsc{FedTimeDis} LSTM algorithm and grouping approach. 
\begin{figure*}[t]
\centering
\includegraphics[width=0.9\textwidth, height=.30\textheight]{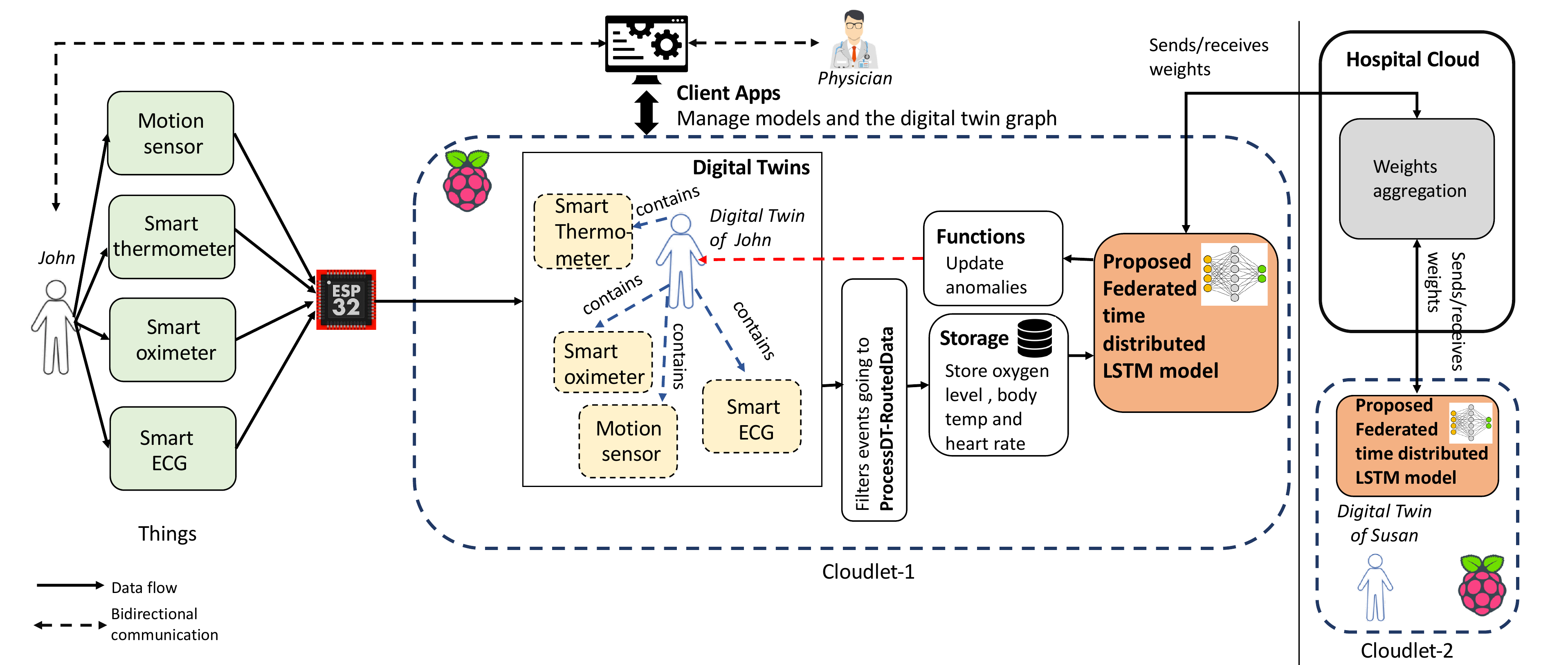}
\centering
\caption{Implementation Framework of RPM Use Case utilizing Digital Twin Service and Edge Cloudlet Computing.}
\label{fig:implement}
\end{figure*}

\subsubsection{Hierarchical Federated Learning Based Anomaly Detection}

The patient's data flows from their DT to local storage for building the proposed AD model. This local model completes training and collaborates with other similar legitimate participants using a grouping approach (discussed in next subsection) in \textit{smart healthcare organization-1}. After updating the local model with global gradients, the training starts again until the minimal error is attained using proposed \textsc{FedTimeDis} LSTM approach. This proposed AD model is also tested using the same approach. 

In this research, a patient's unsafe/anomalous behavior or any attacks on smart healthcare ecosystem can be detected on the edge without mixing up multiple data sources. Since the training data will be fed to the model in real-time, there is less likelihood that the model will be sensitive to small fluctuations as often observed in mass collection of training sets for a series of repeated training and parameter aggregations. This will not lead to overfitting, which usually adversely affects the model performance.
The proposed hierarchical federated learning approach enables multiple health organizations to collaborate with each other.
Figure \ref{fig:usecase} shows that patients \textit{Bob} and \textit{Alice}, who are \textit{OSA} patients of \textit{smart healthcare organization-1}, collaborate together by exchanging their local gradients ($\Delta{w^1}$, $\Delta{w^2}$) with Hospital cloud server-1 to build robust AD model. 
They can also enhance the accuracy of their models by collaborating with other \textit{OSA} patients (\textit{Susan, Max}) of \textit{smart healthcare organization-2} using this proposed hierarchical federated approach. Their global gradients ($\Delta{w_{global1}}$, $\Delta{w_{global3}}$) aggregate at federated multi-cloud server to build a global model $\Delta{w_{fedglobal}}$ at the top layer. 
This approach allows participants to feed local models with new samples, which also helps to reduce false positive rates. For example, the normal range of patient-1's oxygen level is between 95\% to 98\%, and outside of this range is considered as abnormal according to patient-1's device level AD model. On the other side, the normal range of patient-2's oxygen level is between 94\% to 98\%. With the merging of local gradients, it allows to avoid false positives where 94\% oxygen level of patient-1 would be considered normal.

In the other case, the oxygen level and heart rate of a patient have a correlation, such as if the patient's oxygen level drops, then his/her heart rate will also decrease. Some possible attacks including data modification and zero day attack can modify the patient's heart rate. However, anomalies associated with one device (e.g., heart rate sensor) can be detected utilizing strong correlation rules with the patient's other devices and federation with other patient's device data. 
\subsubsection{Grouping Approach}
Figure \ref{fig:usecase} depicts that two groups are formed on the \textit{hospital cloud server-1} based on patients' attributes and their disease names. The grouping of AD models based on specific disease is introduced to categorize patients with similar characteristics into a group. For example, the anomalies of a \textit{OSA} patient are different from a \textit{DB} patient, and the behavioral data of healthy young age patients is different from elderly patients. In order to build groups, we need attributes of patients including age group, disease name, and medications. This approach increases AD model accuracy and reduces training time and response time. The grouping process will occur at the hospital cloud server to classify the AD models based on the patient's attributes. 
The local gradients are sent to the hospital cloud server along with attribute tags, which helps to identify and merge similar groups on hospital cloud server based on patient attributes.    
\subsection{Implementation Framework}
We present our prototype implementation here, which can be considered as a baseline and can be further extended or customized for a large number of smart devices and patients as required. This implementation framework consists of -- smart devices as per our use case requirements, ESP32, Raspberry-Pi, DT service, customized LSTM model and other services, as shown in Figure \ref{fig:implement}. It shows a patient \textit{John}, who uses a set of devices, such as motion sensor, smart thermometer, smart oximeter, and smart ECG at physical device level. We identify these smart sensors as AM312 PIR motion sensor, MAX30100 I2C pulse oximeter sensor and DS18B20 one-wire waterproof temperature sensor, which can be attached to ESP-WROOM-32 through GPIO (general-purpose input/output) pins. The merger of ESP32 and IoT is an innovative technology in the healthcare ecosystem. EPS32 is a powerful microcontroller, which can communicate with other Wi-Fi and Bluetooth devices via its SPI/SDIO, or I2C/UART interfaces. It is small in size and can be designed as a wearable device along with these sensors. First, these sensors sense the raw data of \textit{John} and send it to a Raspberry-Pi through ESP32.
In this research, we envision Raspberry-Pi as ECC node since we can design it to have required computing power and storage needed to run the AD model at low cost. 
For implementing the DT, there are several approaches utilizing existing real-world implementations including Azure Digital Twins, IBM, GE, or we can build our own DT service based on a patient's data. We use Azure DT service\footnote{https://docs.microsoft.com/en-us/azure/digital-twins/overview} to create DT of \textit{John}.

{\small
\begin{verbbox}
{
  "@id": "patient:example:Oximeter;1",
  "@type": "Interface",
  "displayName": "Oximeter",
  "contents": [
    {
      "@type": "Property",
      "name": " blood_oxygen_level",
      "schema": "double"
    },
    {
      "@type": "Property",
      "name": "pulse_rate",
      "schema": "double"
    }
  ],
  "@context": "patient:dtdl:context;2"
}
\end{verbbox}
}
{
\theverbbox
\captionof{Code}{JSON program fragment of Smart Oximeter for Azure Digital Twins.}
}

First, we create a digital twin instance, set up authentication via roles, and feed the DT by IoT and its edge device data using REST APIs. After a DT instance is created, owner/administrator of the RPM ecosystem can assign specific roles to specific users and physicians. Azure DTs use Explorer for visualization, write queries, and edit models and relationships. The models for Azure Digital Twins are written in Digital Twin Definition Language (DTDL), and saved as .json files. In this framework, \textit{John} contains four interfaces, one of them is a smart oximeter interface which implements two telemetry data -- blood oxygen level and pulse rate as shown in Code 1. In the next phase, patient's data in DTs service can be routed towards AD model in cloudlet to detect anomalies. 
The local parameters of the AD model are sent to the hospital cloud server to exchange with other AD models using a grouping approach to enhance the accuracy of local AD models. Our proposed model identifies anomalies and sends alerts to the patient's DT through functions as shown in Figure \ref{fig:implement}. 



\section{Conclusion and Future Work}
\label{conclusion}
IoMT has been propagating in the field of healthcare, and provides better care to the patients at home. For accurate prognosis, an effective AD approach is required to provide patients with relevant information concerning their health.
In this paper, we presented a threat model with specific threat scenarios for centralized AD and proposed a privacy-preserving AD model using \textsc{FedTimeDis} LSTM approach for mitigating those threats. With respect to our research questions, the proposed framework provides: (i) data security and privacy through edge cloudlet deployment of FL based AD model, (ii) model integrity and statistical features through distributed model training and only sharing gradients, (iii) reduced computation and communication costs via local training, and (iv) disease-based grouping and hierarchical FL enable high accuracy.
Our approach utilizes parameter sharing and aggregation as opposed to patient data sharing for detecting anomalous traits. The proposed approach is robust, scalable, and can learn long term dependencies inherent in IoMT sensor data. We further presented a RPM use case and implementation using DT and ECC. Our disease-based grouping of AD models improves model accuracy, and hierarchical FL enables easier collaboration among multiple healthcare organizations. 
Our future work entails data collection and full fledged implementation of our approach along with its performance evaluation. We also plan to demonstrate practical examples of how our proposed approach addresses the limitations of centralized AD. We further plan to evaluate detailed metrics including detection accuracy, false positive rate, recall, F1 score, training time and communication cost of our proposed model.

 {
    \bibliographystyle{plain}
    \bibliography{References}
    }
\end{document}